\documentstyle[aps,prl]{revtex}
\newcommand{\BE}{\begin{equation}}
\newcommand{\EE}{\end{equation}}
\newcommand{\BA}{\begin{eqnarray}}
\newcommand{\EA}{\end{eqnarray}}

\begin{document}
\draft

\twocolumn[\hsize\textwidth\columnwidth\hsize\csname@twocolumnfalse\endcsname
\title{Comment on ``Statistics of the Lyapunov Exponent in 1D Random periodic-on-Average Systems"
[Phys. Rev. Lett. {\bf 81}, 5390, 1998]}
\author{P.-G. Luan and Z. Ye}
\address{Wave Phenomena Laboratory, Department of Physics, National Central University, Chung-li, Taiwan 320}
\date{\today}
\maketitle

]

In this comment, we would like to point out possible numerical
errors in a recent Letter by Deych {\it et al.}\cite{Deych}.

In \cite{Deych}, the authors pointed out an interesting
observation that in some 1D random periodic-on-average systems,
states from pass bands and band edges of the underlying band
structure follow single parameter scaling with universal behavior,
whereas states from the interior of the band gaps do not have
universal behavior and require two parameters to describe their
scaling properties. When the degree of disorder exceeds a certain
threshold the single parameter scaling is restored for an entire
band gap. Upon careful inspection, we find that while these
conclusions are qualitatively valid, there may be some numerical
errors in the results of \cite{Deych}.

\input{epsf}
\begin{figure}[hbt]
\begin{center}
\epsfxsize=2.in \epsffile{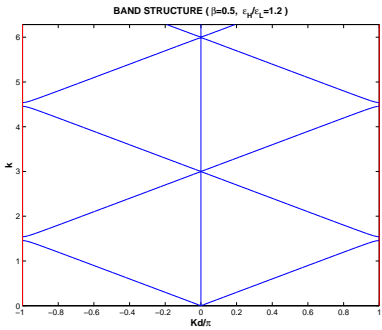} \vspace{10pt}
\caption{\label{figure1}\small Band structure}
\end{center}
\end{figure}

The model and all the parameters used in the following computation
are the same as in \cite{Deych}. The model in \cite{Deych} is
essentially the same as the second model in \cite{McGurn}. First
we compute the band structure using Eq.~(6) of \cite{McGurn} the
result is presented in Fig.~1. It shows that the first band gap
occurs between $k=1.456$ and $1.543$ and there is a very narrow
second gap located between $k=2.992$ and $3.005$. The third band
gap is at around $k=4.5$. In \cite{Deych}, however, the results
indicates a band gap at around $k=2.86$ which is absent in our
computation. To validate our program, we used the parameters from
\cite{McGurn} in the computation and obtained the same band
structure results as in \cite{McGurn}. Moreover, by adjusting the
parameters we were also able to reproduce almost exactly the same
result for the disorder dependence of the Lyapunov exponent as in
Fig.~1 of \cite{Maradudin}.

Following \cite{Deych}, we then compute the disorder dependence of
the Lyapunov exponent on the standard deviation in the randomness
of the layers' thickness denoted by $\sigma$, and the frequency
dependence of the Lyapunov exponent and its variance for
frequencies covering the first three band gaps. The results are
shown in Fig.~2. It is clear that the results for frequencies
covering the first band gaps are qualitatively in agreement with
the results in Fig.~1 and Fig.~2 of \cite{Deych}, except that the
band gap in \cite{Deych} is implied at around $k=2.86$. In
particular, as the randomness increases, all the Lyapunov exponent
curves merge. For frequencies in the vicinity of the second band
gap, however, the exponent curves come across into each other at a
certain level of disorder, but split again as the amount of
disorder increases further. This is shown in Fig.~2(a2). In this
case, for the same level of the disorder the double peaked
structure of the variance observed in the first band gap
disappears (Fig.~1(b2)). When the disorder decreases, to $\sigma
=0.00577$ say, the double peaks appear again. For frequencies near
the third band gap, as indicated by Fig.~2(a3) and (b3) the
results appear to follow the features at frequencies around the
first band gap. For comparison, we also plot in Fig.~3 the
dependence of the Lyapunov exponent on the standard deviation
$\sigma$ of the layer randomness at the frequencies indicated in
Fig.~2 of \cite{Deych}. At these frequencies, all curves seem to
merge.

\input{epsf}
\begin{figure}[hbt]
\begin{center}
\epsfxsize=2.95in \epsffile{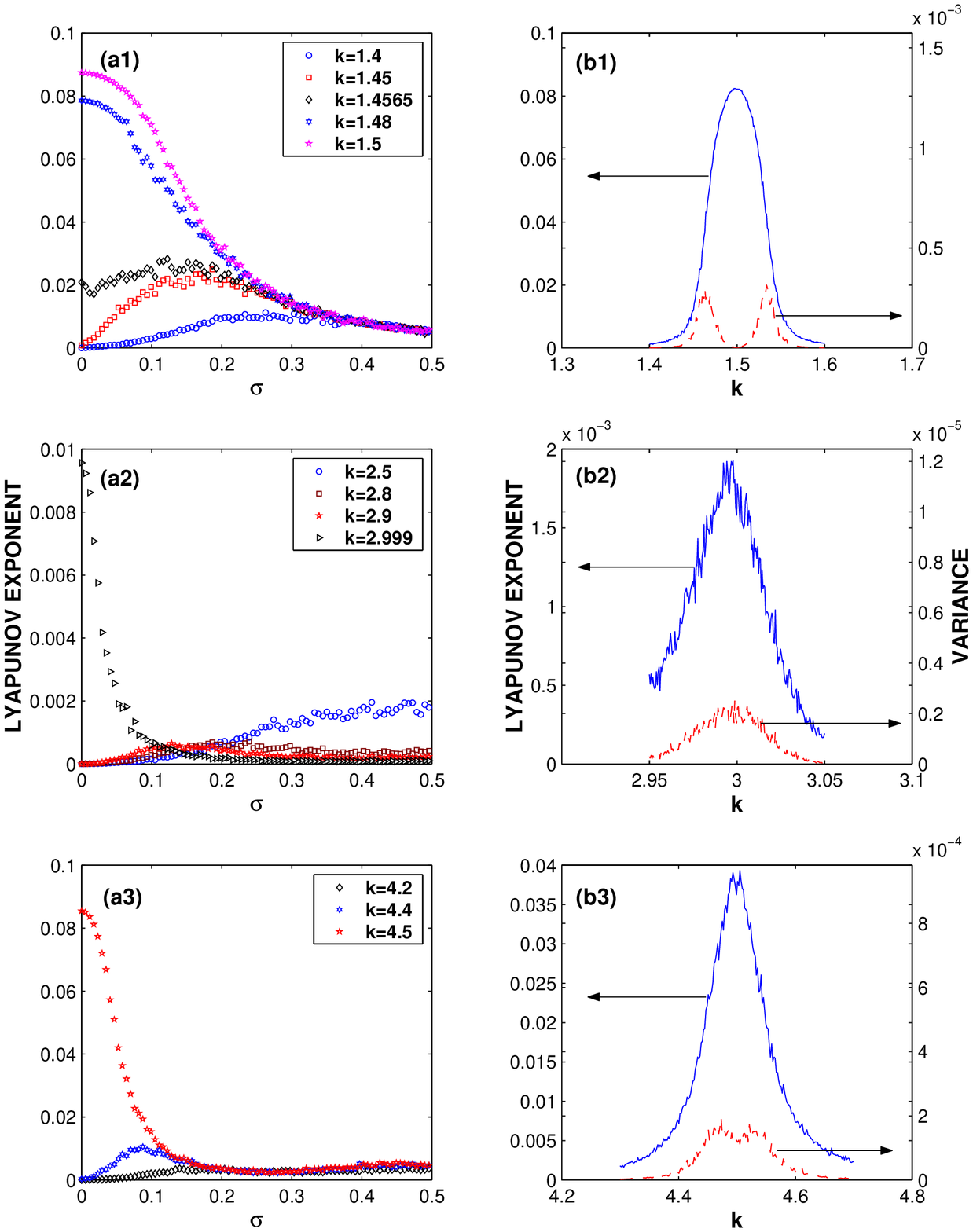} \vspace{10pt}
\caption{\label{figure2}\small Left panel: The dependence of the
Lyapunov exponent on for different $k$ in the around the first,
second and third band gaps respectively. Right panel: Frequency
dependence of the Lyapunov exponent and its variance for the
frequencies covering the first, second and the third band gaps.
For the right panel, $\sigma$ is taken as $0.0577$.}
\end{center}
\end{figure}

\input{epsf}
\begin{figure}[hbt]
\begin{center}
\epsfxsize=2.5in \epsffile{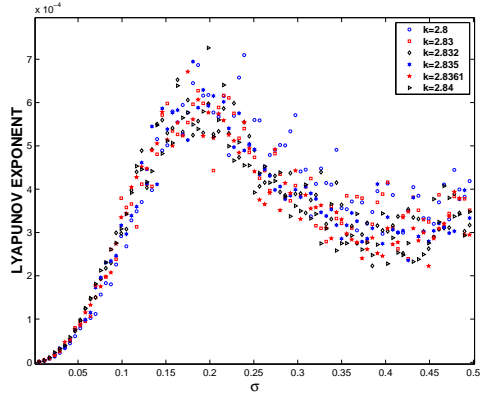} \vspace{10pt}
\caption{\label{figure3}\small The dependence of the Lyapunov
exponent on the standard deviation $\sigma$ of the random layers'
thickness for different $k$.}
\end{center}
\end{figure}

This work received support from National Science Council (Grant
No. NSC-89-2112-M008-008 and NSC-89-2611-M008-002).

\end{document}